\documentclass[a4paper,12pt]{article}
\usepackage[latin1]{inputenc}
\usepackage{amssymb}

\title{{\bf  An Upper Bound on the Complexity  \\ of  Recognizable Tree Languages   }}

\author{Olivier Finkel 
\\ {\it Equipe de Logique Math\'ematique }  
\\ Institut de Math\'ematiques de Jussieu - Paris Rive Gauche 
\\ CNRS et  Universit\'e Paris Diderot Paris 7
\\ B\^{a}timent Sophie Germain Case 7012 
\\ 75205 Paris Cedex 13, France
\\ E Mail: finkel@math.univ-paris-diderot.fr
\vspace{08mm}
\\ Dominique Lecomte 
\\ {\it Projet Analyse Fonctionnelle } 
\\ Institut de Math\'ematiques de Jussieu - Paris Rive Gauche 
\\ Universit\'e Pierre et Marie Curie Paris 6
\\ Couloir 16-26, 4\`eme \'etage, Case 247, 4, place Jussieu, 
\\ 75252 Paris Cedex 05, France
\\ E Mail: dominique.lecomte@upmc.fr
\\ \\ Universit\'e de Picardie, I.U.T. de l'Oise, site de Creil
\\ 13, all\'ee de la fa\"iencerie, 
\\ 60107 Creil, France
\vspace{08mm}
\\ Pierre Simonnet 
\\ {\it UMR CNRS 6134} 
\\ Facult\'e des Sciences
\\ Universit\'e de Corse
\\ Quartier Grossetti BP52 20250, Corte, France
\\ E Mail: simonnet@univ-corse.fr} 

\date{}

\begin{document}

\newtheorem{The}{Theorem}[section]
\newtheorem{Pro}[The]{Proposition}
\newtheorem{Deff}[The]{Definition}
\newtheorem{Lem}[The]{Lemma}
\newtheorem{Rem}[The]{Remark}
\newtheorem{Exa}[The]{Example}
\newtheorem{Cor}[The]{Corollary}

\newcommand{\fa}{\forall}
\newcommand{\Ga}{\Gamma}
\newcommand{\Gas}{\Gamma^\star}
\newcommand{\Gao}{\Gamma^\omega}

\newcommand{\Si}{\Sigma}
\newcommand{\Sis}{\Sigma^\star}
\newcommand{\Sio}{\Sigma^\omega}
\newcommand{\ra}{\rightarrow}
\newcommand{\hs}{\hspace{12mm}

}
\newcommand{\lra}{\leftrightarrow}
\newcommand{\la}{language}
\newcommand{\ite}{\item}
\newcommand{\Lp}{L(\varphi)}
\newcommand{\abs}{\{a, b\}^\star}
\newcommand{\abcs}{\{a, b, c \}^\star}
\newcommand{\ol}{ $\omega$-language}
\newcommand{\orl}{ $\omega$-regular language}
\newcommand{\om}{\omega}
\newcommand{\nl}{\newline}
\newcommand{\noi}{\noindent}
\newcommand{\tla}{\twoheadleftarrow}
\newcommand{\de}{deterministic }
\newcommand{\proo}{\noi {\bf Proof.} }
\newcommand {\ep}{\hfill $\square$}
\renewcommand{\thefootnote}{\star{footnote}} 

\newcommand{\bormxi}{{\bf\Pi}^{0}_{\xi}}
\newcommand{\bormlxi}{{\bf\Pi}^{0}_{<\xi}}
\newcommand{\bormz}{{\bf\Pi}^{0}_{0}}
\newcommand{\bormone}{{\bf\Pi}^{0}_{1}}
\newcommand{\ca}{{\bf\Pi}^{1}_{1}}
\newcommand{\bormtwo}{{\bf\Pi}^{0}_{2}}
\newcommand{\bormthree}{{\bf\Pi}^{0}_{3}}
\newcommand{\bormom}{{\bf\Pi}^{0}_{\omega}}
\newcommand{\borml}{{\bf\Pi}^{0}_{\lambda}}
\newcommand{\bormlpn}{{\bf\Pi}^{0}_{\lambda +n}}
\newcommand{\bormpm}{{\bf\Pi}^{0}_{1+m}}
\newcommand{\bormep}{{\bf\Pi}^{0}_{\eta +1}}
\newcommand{\borapxi}{{\bf\Sigma}^{0}_{1+\xi}}
\newcommand{\borai}{{\bf\Sigma}^{0}_{ 2.\xi +1 }}
\newcommand{\bormpxi}{{\bf\Pi}^{0}_{1+\xi}}
\newcommand{\bormpeta}{{\bf\Pi}^{0}_{1+\eta}}
\newcommand{\bormeta}{{\bf\Pi}^{0}_{\eta}}
\newcommand{\borapxipo}{{\bf\Sigma}^{0}_{1+\xi +1}}
\newcommand{\bormpxipo}{{\bf\Pi}^{0}_{1+\xi +1}}
\newcommand{\bormxipo}{{\bf\Pi}^{0}_{\xi +1}}
\newcommand{\borpxi}{{\bf\Delta}^{0}_{1+\xi}}
\newcommand{\borel}{{\bf\Delta}^{1}_{1}}
\newcommand{\Borel}{{\it\Delta}^{1}_{1}}
\newcommand{\borone}{{\bf\Delta}^{0}_{1}}
\newcommand{\bortwo}{{\bf\Delta}^{0}_{2}}
\newcommand{\borthree}{{\bf\Delta}^{0}_{3}}
\newcommand{\boraone}{{\bf\Sigma}^{0}_{1}}
\newcommand{\boratwo}{{\bf\Sigma}^{0}_{2}}
\newcommand{\borathree}{{\bf\Sigma}^{0}_{3}}
\newcommand{\boraom}{{\bf\Sigma}^{0}_{\omega}}
\newcommand{\boraxi}{{\bf\Sigma}^{0}_{\xi}}
\newcommand{\ana}{{\bf\Sigma}^{1}_{1}}
\newcommand{\pca}{{\bf\Sigma}^{1}_{2}}
\newcommand{\Ana}{{\it\Sigma}^{1}_{1}}
\newcommand{\Boraone}{{\it\Sigma}^{0}_{1}}
\newcommand{\Borone}{{\it\Delta}^{0}_{1}}
\newcommand{\Bormone}{{\it\Pi}^{0}_{1}}
\newcommand{\Bormtwo}{{\it\Pi}^{0}_{2}}
\newcommand{\Ca}{{\it\Pi}^{1}_{1}}
\newcommand{\bormn}{{\bf\Pi}^{0}_{n}}
\newcommand{\bormm}{{\bf\Pi}^{0}_{m}}
\newcommand{\boralp}{{\bf\Sigma}^{0}_{\lambda +1}}
\newcommand{\borat}{{\bf\Sigma}^{0}_{|\theta |}}
\newcommand{\bormat}{{\bf\Pi}^{0}_{|\theta |}}
\newcommand{\Borapxi}{{\it\Sigma}^{0}_{1+\xi}}
\newcommand{\Boraxi}{{\it\Sigma}^{0}_{\xi}}
\newcommand{\Bormxi}{{\it\Pi}^{0}_{\xi}}
\newcommand{\Bormxipo}{{\it\Pi}^{0}_{\xi +1}}
\newcommand{\Bormpxipo}{{\it\Pi}^{0}_{1+\xi +1}}
\newcommand{\Borapn}{{\it\Sigma}^{0}_{1+n}}
\newcommand{\Boran}{{\it\Sigma}^{0}_{n}}
\newcommand{\borapn}{{\bf\Sigma}^{0}_{1+n}}
\newcommand{\boran}{{\bf\Sigma}^{0}_{n}}
\newcommand{\boraxim}{{\bf\Sigma}^{0}_{\xi^-}}
\newcommand{\Boratwo}{{\it\Sigma}^{0}_{2}}
\newcommand{\Borathree}{{\it\Sigma}^{0}_{3}}
\newcommand{\Borthree}{{\it\Delta}^{0}_{3}}
\newcommand{\Borapnpo}{{\it\Sigma}^{0}_{1+n+1}}
\newcommand{\Boranpo}{{\it\Sigma}^{0}_{n+1}}
\newcommand{\Bormpxi}{{\it\Pi}^{0}_{1+\xi}}
\newcommand{\Borpxi}{{\it\Delta}^{0}_{1+\xi}}
\newcommand{\Borxi}{{\it\Delta}^{0}_{\xi}}
\newcommand{\borxi}{{\bf\Delta}^{0}_{\xi}}
\newcommand{\Bormlxi}{{\it\Pi}^{0}_{<\xi}}
\newcommand{\borapeap}{{\bf\Sigma}^{0}_{1+\eta_{\alpha ,p}}}
\newcommand{\borapeapn}{{\bf\Sigma}^{0}_{1+\eta_{\alpha ,p,n}}}
\newcommand{\Borapeap}{{\it\Sigma}^{0}_{1+\eta_{\alpha ,p}}}
\newcommand{\Bormpn}{{\it\Pi}^{0}_{1+n}}
\newcommand{\Bormn}{{\it\Pi}^{0}_{n}}
\newcommand{\Borpn}{{\it\Delta}^{0}_{1+n}}
\newcommand{\Born}{{\it\Delta}^{0}_{n}}
\newcommand{\borapximo}{{\bf\Sigma}^{0}_{1+(\xi -1)}}
\newcommand{\boreta}{{\bf\Delta}^{0}_{\eta}}
\newcommand{\pn}{{\bf\Delta}^{1}_{n}}
\newcommand{\pan}{{\bf\Sigma}^{1}_{n}}
\newcommand{\pmn}{{\bf\Pi}^{1}_{n}}
\newcommand{\panpo}{{\bf\Sigma}^{1}_{n+1}}

\maketitle

\begin{abstract}
\noi  The third author noticed in his 1992 PhD Thesis \cite{Simonnet92} that every regular tree language of infinite trees is in a class $\Game (D_n({\bf \Sigma}_2^0))$ for some natural number $n\geq 1$, where $\Game$ is the game quantifier. We first give a detailed exposition of this result. Next, using an embedding of the Wadge hierarchy of non self-dual Borel subsets of the Cantor space $2^\om$ into the class ${\bf \Delta}_2^1$, and the notions of Wadge degree and Veblen function, we argue that this upper bound on the topological complexity of regular tree languages is much better than the usual  ${\bf \Delta}_2^1$.  
\end{abstract}

\hs \noi {\bf keywords:}  
infinite trees; tree automaton; regular tree language; Cantor topology; topological complexity; 
Borel hierarchy; game quantifier; Wadge classes; Wadge degrees; universal sets; 
provably-$\Delta_2^1$. 

\section{$\!\!\!\!\!\!\!$ Introduction}\indent

 A way to study the complexity of languages of infinite words or infinite trees accepted by various kinds of automata is to study their topological complexity,  \cite{LescowThomas,PerrinPin,2001automata,Staiger97,Thomas90,ADMN}. In this paper we consider the topological complexity of regular languages of trees.\bigskip 

 On one side the topological complexity of {\it deterministic } regular languages of trees has been determined. Niwinski and Walukiewicz proved that a tree language which is accepted by a 
deterministic Muller  tree automaton is either in the class ${\bf \Pi}^0_3$ or ${\bf \Pi}^1_1$-complete,    \cite{NiwiskiWalukiewicz03}. And the Wadge hierarchy of {\it deterministic } regular languages of trees has been determined by Murlak, \cite{Murlak-LMCS, ADMN}.\bigskip

 On the other side in the course of years, more and more complex (non-determinis-tic) 
regular languages of trees have been found. Skurczynski proved that, for every natural number 
$n\geq 1$, there are some ${\bf \Pi}^0_n$-complete and some ${\bf \Si}^0_n$-complete regular tree languages, \cite{Skurczynski93}.  Notice that it is an  open question to know whether there exist some regular sets of trees which are Borel sets of infinite rank. Niwinski showed that there are some ${\bf \Si}^1_1$-complete regular sets of trees accepted by B\"uchi tree automata, and some ${\bf \Pi}^1_1$-complete regular sets of trees accepted by deterministic Muller  tree automata, 
\cite{Niwinski85}.  
 Some  examples of regular  tree  languages 
 at some transfinite levels of the  difference hierarchy of analytic sets 
 were given by Simonnet and Finkel in \cite{Simonnet92,Fink-Sim-trees}. 
Hummel proved in \cite{Hummel12} that there exists some (unambiguous) regular tree language which is topologically more complex than any set in the difference hierarchy of analytic sets. Arnold and Niwinski showed in \cite{ArnoldNiwinski08} that the game tree languages  $W_{(\iota, \kappa)}$ form a infinite hierarchy of non Borel regular sets of trees with regard to the Wadge reducibility.

\vfill\eject

 An upper bound on the  complexity of regular languages of trees follows from the definition of acceptance by  non deterministic Muller or Rabin automata and from Rabin's complementation Theorem: every regular set of trees is a ${\bf \Delta}^1_2$-set, see \cite{Rabin69,PerrinPin,Thomas90,LescowThomas}.\bigskip

The third author noticed in his 1992 PhD Thesis \cite{Simonnet92} that every regular tree language of infinite trees is in a class $\Game (D_n({\bf \Sigma}_2^0))$ for some natural number $n\geq 1$, where $\Game$ is the game quantifier. We first give a detailed exposition of this result. Next, using an embedding of the Wadge hierarchy of non self-dual Borel subsets of the Cantor space $2^\om$ into the class ${\bf \Delta}_2^1$, and the notions of Wadge degree and Veblen function, we argue that this upper bound on the topological complexity of regular tree languages is much better than the usual  ${\bf \Delta}_2^1$.\bigskip  

 The paper is organized as follows. In Section 2 we recall the notions of Muller tree automaton and regular tree language. The notions of topology are recalled in Section 3.  We give the upper bound on the complexity of regular tree languages in Section  4. We argue that it is a much better upper bound than ${\bf \Delta}^1_2$ in Section 5.

\section{$\!\!\!\!\!\!\!$ Recognizable tree languages}\indent

 We now recall usual notation of formal language theory.\bigskip
  
\noindent $\bullet$ In the sequel, $\Si$ will be a finite alphabet with at least two letters. A 
{\bf non-empty finite word} over $\Si$ is a sequence $x=a_1\cdots a_k$, where $a_i\in\Sigma$ for $i=1,\ldots ,k$, and  $k\geq 1$ is a natural number. The {\bf length} of $x$ is $k$, and denoted by 
$|x|$. The {\bf empty word} has no letter and is denoted by $\lambda$; its length is $0$. $\Sis$  is the {\bf set of finite words} (including the empty word) over $\Sigma$. A {\bf finitary language} over 
$\Si$ is a subset of $\Sis$.\bigskip 

\noindent $\bullet$ The {\bf first infinite ordinal} is $\om$. An $\om$-{\bf word} over $\Si$ is an 
$\om$-sequence $a_1 \cdots a_n \cdots$, where $a_i\in\Sigma$ for each natural number 
$i\geq 1$.  When $\sigma$ is an $\om$-word over $\Si$, we write 
$\sigma =\sigma(1)\sigma(2)\cdots \sigma(n) \cdots$, where $\sigma(i)\in \Si$ for each $i$, 
$\sigma[n]=\sigma(1)\sigma(2)\cdots \sigma(n)$ for each $n\geq 1$, and $\sigma[0]=\lambda$. The {\bf set of} $\om$-{\bf words} over $\Si$ is denoted by $\Si^\om$. An $\om$-{\bf language} over 
$\Sigma$ is a subset of $\Si^\om$.\bigskip 

\noindent $\bullet$ The usual concatenation product of two finite words $u$ and $v$ is denoted by 
$u\cdot v$ (and sometimes just $uv$). This product is extended to the product of a finite word $u$ and an $\om$-word $v$: the infinite word $u\cdot v$ is then the $\om$-word such that: 
$(u\cdot v)(k)=u(k)$ if $k\leq |u|$, and $(u\cdot v)(k)=v(k-|u|)$ if $k>|u|$.\bigskip

\noindent $\bullet$ The {\bf prefix relation} is denoted by $\sqsubseteq$: a finite word $u$ is a 
{\bf prefix} of a finite word $v$ (respectively,  an infinite word $v$), denoted $u\sqsubseteq v$, if and only if there exists a finite word $w$ (respectively,  an infinite word $w$), such that $v=u\cdot w$.\bigskip

\noindent $\bullet$ We now introduce the languages of infinite binary trees whose nodes are labelled in a finite alphabet $\Si$. A node of an infinite binary tree is represented by a finite  word over the alphabet $\{l, r\}$, where $l$ means ``left" and $r$ means ``right". An infinite binary tree whose nodes are labelled  in $\Si$ is identified with a function $t: \{l, r\}^\star \ra \Si$. The 
{\bf set of  infinite binary trees labelled in} $\Si$ will be denoted by $T_\Si^\om$.\bigskip

\noindent $\bullet$ Let $t$ be a tree. A {\bf branch} $b$ of $t$ is a subset of the set of nodes of $t$ which is linearly ordered by the tree partial order $\sqsubseteq$ and closed under prefix relation, i.e., if $x$ and $y$ are nodes of $t$ such that $y\in b$ and $x \sqsubseteq y$, then $x\in b$. A branch $b$ of a tree is said to be {\bf maximal} if there is no branch of $t$ containing strictly $b$.\bigskip

 Let $t$ be an infinite binary tree in $T_\Si^\om$. If $b$ is a maximal branch of $t$, then this branch is infinite. Let $(x_i)_{i\geq 0}$ be the enumeration of the nodes in $b$ which is strictly increasing for the prefix order. The infinite sequence of labels of the nodes of  such a maximal branch $b$, 
 i.e., $t(x_0)t(x_1) \cdots t(x_n) \cdots $, is called a {\bf path}. It is an $\om$-word over $\Si$.\bigskip

 We now  define tree automata and regular tree languages. 

\begin{Deff} A (nondeterministic topdown) {\bf tree automaton} is a quadruple 
$\mathcal{A}=(Q,\Si,\Delta, q_0)$, where $Q$ is a finite set of states, $\Sigma$ is a finite input alphabet, $q_0 \in Q$ is the initial state and $\Delta \subseteq Q\times\Si\times Q\times Q$ is the transition relation. The tree automaton  $\mathcal{A}$ is said to be {\bf deterministic} if the relation 
$\Delta$ is a functional one, i.e., if for each $(q, a)\in Q\times\Si$ there is at most one pair of states 
$(q', q'')$ such that  $(q, a, q', q'') \in \Delta$. A {\bf run} of the tree automaton 
$\mathcal{A}$ on an infinite binary tree $t\in T_\Si^\om$ is a infinite binary tree $\rho \in T_Q^\om$ such that $\rho (\lambda)=q_0$ and, for each $x\in\{l, r\}^\star$, 
$(\rho(x), t(x), \rho(x\cdot l), \rho(x\cdot r))\in\Delta$.\end{Deff}

\begin{Deff} A {\bf Muller} (nondeterministic topdown) {\bf tree automaton} is a 5-tuple 
$\mathcal{A}=(Q,\Si,\Delta, q_0, \mathcal{F})$, where $(Q,\Si,\Delta, q_0)$ is a tree automaton and $\mathcal{F} \subseteq 2^Q$ is the collection of  designated state sets. A run $\rho$ of the  Muller  tree automaton $\mathcal{A}$ on an infinite binary tree $t\in T_\Si^\om$ is said to be {\bf accepting} if, for each path $p$ of $\rho$, the set of states appearing infinitely often in this path is in 
$\mathcal{F}$. The {\bf tree language} $L(\mathcal{A})$ {\bf accepted by the Muller tree automaton} $\mathcal{A}$ is the set of infinite binary trees $t\in T_\Si^\om$ such that there is (at least) one accepting run of $\mathcal{A}$ on $t$. The class {\bf $REG$} of {\bf regular}, or recognizable, 
{\bf tree languages} is the class of  tree  languages accepted by some Muller automaton.\end{Deff}

\begin{Rem} A tree language is accepted by a Muller tree  automaton iff it is accepted by some Rabin tree automaton iff it is accepted by some parity tree automaton. We refer for instance to 
\cite{Thomas90,2001automata,PerrinPin,ADMN} for the definition of Rabin tree  automaton and of parity tree automaton.\end{Rem}

\section{$\!\!\!\!\!\!\!$ Topology}\indent

 We assume the reader to be familiar with basic notions of topology which may be found in 
\cite{Moschovakis09,LescowThomas,Kechris94,Staiger97,PerrinPin}. There is a natural metric on the set $\Sio$, which is called the {\bf prefix metric} and defined as follows. For 
$\sigma ,\sigma'\in \Sio$ and $\sigma\neq\sigma'$ let 
$\delta(\sigma ,\sigma')=2^{-l_{\mathrm{pref}(\sigma ,\sigma')}}$ where 
$l_{\mathrm{pref}(\sigma ,\sigma')}$ is the first natural number $n$ such that the $(n+1)^{st}$ letter of $\sigma$ is different from the $(n+1)^{st}$ letter of $\sigma'$. This metric induces on $\Sio$ the usual topology for which {\bf open} subsets of $\Sio$ are of the form $W\cdot \Si^\om$, where 
$W\subseteq \Sis$. A set $L\subseteq \Si^\om$ is {\bf closed} if its complement 
$\Si^\om\setminus L$ is an open set. The topological space $\Sio$ is a Cantor space equipped with the usual Cantor topology.\bigskip 

\noindent $\bullet$ There is also a natural topology on the set $T_\Si^\om$ 
\cite{Moschovakis09,LescowThomas,Kechris94}. It is defined by the following distance. Let $t$ and $s$ be two distinct infinite trees in $T_\Si^\om$. Then the distance between $t$ and $s$ is 
$\frac{1}{2^n}$ where $n$ is the smallest natural number such that $t(x)\neq s(x)$ for some word 
$x\in \{l, r\}^\star$ of length $n$. The open sets are then in the form $T_0\cdot T_\Si^\om$ where $T_0$ is a set of finite labelled trees. $T_0\cdot T_\Si^\om$ is the set of infinite binary trees 
which extend some finite labelled binary tree $t_0\in T_0$, $t_0$ is here a sort of prefix, 
an ``initial subtree" of a tree in $t_0\cdot T_\Si^\om$. It is well known that the set $T_\Si^\om$, equipped with this topology, is homeomorphic to the Cantor space $2^\om$, hence also to the topological spaces $\Sio$, where $\Si$ is finite.\bigskip 

\noindent $\bullet$ We  now define the Borel Hierarchy of subsets of $\Si^\om$. It is defined similarly on the space $T_\Si^\om$.

\begin{Deff} For a countable ordinal $\xi\geq 1$, the classes ${\bf \Si}^0_\xi$ and ${\bf \Pi}^0_\xi$ of the {\bf Borel Hierarchy} on the topological space $\Si^\om$ are defined as follows:\smallskip

\noindent ${\bf \Si}^0_1$ is the class of open subsets of $\Si^\om$,\smallskip

\noindent ${\bf \Pi}^0_1$ is the class of closed subsets of $\Si^\om$,\smallskip

\noindent and, for any countable ordinal $\xi\geq 2$,\smallskip

\noindent ${\bf \Si}^0_\xi$ is the class of countable unions of subsets of $\Si^\om$ in 
$\bigcup_{\eta <\xi}{\bf \Pi}^0_\eta$,\smallskip

\noindent ${\bf \Pi}^0_\xi$ is the class of countable intersections of subsets of $\Si^\om$ in 
$\bigcup_{\eta <\xi}{\bf \Si}^0_\eta$.\end{Deff}

 For a countable ordinal $\xi$, a subset of $\Si^\om$ is a Borel set of {\bf rank} $\xi$ if 
it is in ${\bf \Si}^0_{\xi}\cup {\bf \Pi}^0_{\xi}$ but not in 
$\bigcup_{\eta <\xi}({\bf \Si}^0_\eta\cup {\bf \Pi}^0_\eta )$.\bigskip

\noindent $\bullet$ There exists another hierarchy beyond the Borel hierarchy, which is called the 
{\bf projective hierarchy}. The classes ${\bf \Si}^1_n$ and ${\bf \Pi}^1_n$ of the projective hierarchy, defined for natural numbers $n\geq 1$, are obtained from the Borel hierarchy by successive applications of the operations of projection and complementation. The first level of the projective hierarchy is formed by the class ${\bf \Si}^1_1$ of analytic sets and the class ${\bf \Pi}^1_1$ of 
{\bf co-analytic sets}, which are the complements of analytic sets.

\vfill\eject

 In particular, the class of Borel subsets of $\Si^\om$ is strictly contained in the class of analytic sets, which are obtained by projection of Borel sets.

\begin{Deff} A subset $A$ of $\Si^\om$ is {\bf analytic} if there is a Borel subset $B$ of 
$(\Si \times 2)^\om$, where $2=\{0, 1\}$,  such that $\sigma\in A\Leftrightarrow\exists\theta\in 2^\om$ with 
$(\sigma ,\theta ) \in B$, where $(\sigma ,\theta )$ is the infinite word over the alphabet 
$\Si\times 2$ such that $(\sigma ,\theta )(i)=(\sigma (i),\theta (i))$ for each natural number 
$i\geq 1$.\end{Deff} 

\begin{Rem} In the above definition we could take $B$ in the class ${\bf \Pi}^0_2$, 
\cite{Moschovakis09}.\end{Rem}

 The Borel hierarchy and the projective hierarchy on $T_\Si^\om$ are defined from open 
sets in the same manner as in the case of the topological space $\Si^\om$.\bigskip

\noindent $\bullet$ The notion of Wadge reducibility is defined via the reduction by continuous functions. In the sequel, $\Theta$ will be a finite alphabet with at least two letters. For 
$L\subseteq\Si^\om$ and $L'\subseteq\Theta^\om$, $L$ is said to be {\bf Wadge reducible} to $L'$, denoted by $L\leq _W L'$, if there exists a continuous function $f:\Si^\om\ra\Theta^\om$, such that 
$L=f^{-1}(L')$. If $\xi\geq 1$ is a countable ordinal and $n\geq 1$ is a natural number, then a set 
$F\subseteq \Si^\om$ is said to be ${\bf \Si}^0_\xi$ (respectively,  ${\bf \Pi}^0_\xi$, 
${\bf \Si}^1_n$, ${\bf \Pi}^1_n$)-{\bf complete} if, for any set $E\subseteq\Theta^\om$, 
$E\in {\bf \Si}^0_\xi$ (respectively, $E\in {\bf \Pi}^0_\xi$, $E\in {\bf \Si}^1_n$, 
$E\in {\bf \Pi}^1_n$) iff $E \leq_W F$.\bigskip
  
 The $\om$-language $\mathcal{R}=(0^\star\cdot 1)^\om$ is a well known example of a 
${\bf \Pi}^0_2 $-complete subset of $\{0, 1\}^\om$. It is the set of $\om$-words over $\{0, 1\}$ having infinitely many occurrences of the letter $1$. Its  complement 
$\{0, 1\}^\om\setminus (0^\star\cdot 1)^\om$ is a ${\bf \Si}^0_2 $-complete subset of $\{0, 1\}^\om$. The set of infinite trees in $T_\Si^\om$, where $\Si=\{0, 1\}$, having at least one path in the $\om$-language $\mathcal{R}$ is ${\bf \Si}^1_1$-complete.\bigskip

\noindent $\bullet$ We now define the difference hierarchy over a class ${\bf \Gamma}$. Let 
$n$ be a natural number, and $(A_p)_{p<n}$ be an increasing sequence of subsets of some space $Z$. The set $D_n[(A_p)_{p<n}]$ is the set of elements $z\!\in\! Z$ such that 
$z\!\in\! A_p\!\setminus\!\bigcup_{q<p}\ A_q$ for some $p\! <\! n$ whose parity is opposite to that of $n$. We can now define the class of $n$-differences of ${\bf \Gamma}$-subsets of $Z$, where 
$Z=\Si^\om$ or $Z=T_\Si^\om$: 
$$D_n({\bf \Gamma})\! :=\!\{ D_n[(A_p)_{p<n}]\mid A_p\mbox{ is in the class }{\bf\Gamma}
\mbox{ for each }p<n\}.$$
It is well known that, for every countable ordinal $\xi\geq 1$, the hierarchy of differences of 
${\bf \Si}^0_\xi$-sets is strict, i.e., the inclusion $D_m({\bf \Si}^0_\xi)  \subset   D_n({\bf \Si}^0_\xi )$ holds if $m<n$.\bigskip

Notice that the difference hierarchy has an extension to countable ordinal ranks, see for instance \cite{Moschovakis09,Kechris94}, but we shall not need this in the sequel. 
\bigskip

 Moreover, every regular $\om$-language is a boolean combination of ${\bf \Si}^0_2$-sets, and belongs to some class $D_n ({\bf  \Si}^0_2)$, for some natural number $n\geq 1$. We shall also consider these classes $D_n ({\bf  \Si}^0_2)$ in the sequel.\bigskip 

 It is also well known that the hierarchy of differences of  analytic sets is strict. It was proved by Hummel in \cite{Hummel12} that there exists some (unambiguous) regular tree language which is topologically more complex than any set in the difference hierarchy of analytic sets because there exists  some  (unambiguous) regular tree language which does not belong to the $\sigma$-algebra generated by the analytic sets.

\section{$\!\!\!\!\!\!\!$ An  Upper Bound}\indent

 We first recall the definition of Gale-Stewart games. 

\begin{Deff}[\cite{Kechris94}] Let $A\subseteq\Si^\om$. The {\bf Gale-Stewart  game} $G(A)$ is a game with perfect information between two players. Player 1 first writes a letter $a_1\in\Si$, then Player 2 writes a letter $b_1\in\Si$, then Player 1 writes $a_2\in\Si$, and so on $\ldots$ After $\om$ steps, the two players have composed a word $\sigma =a_1b_1a_2b_2\ldots$ of $\Si^\om$. Player 1 wins the play iff 
$\sigma\in A$, otherwise Player 2 wins the play.\smallskip

 Let $A\subseteq\Si^\om$ and $G(A)$ be the associated Gale-Stewart  game. A {\bf strategy} for Player 1 is a function $F_1: (\Si^2)^\star\ra\Si$ and a strategy for Player 2 is a function 
$F_2: (\Si^2)^\star\Si\ra\Si$. Player 1 follows the strategy $F_1$ in a play if  for each natural number $n\geq 1$ ~~  $a_n = F_1(a_1b_1a_2b_2 \cdots a_{n-1}b_{n-1})$. If Player 1 wins every play in which she has followed the strategy $F_1$, then we say that the strategy $F_1$ is a 
{\bf winning strategy} (w.s.)  for Player 1.  The notion of winning strategy for Player 2 is defined in a similar manner.\smallskip  

 The game $G(A)$  is said to be {\bf determined} if one of the two players has a winning strategy.
\end{Deff}


 We now recall the definition of the {\bf game quantifier}. 

\begin{Deff}[\cite{Moschovakis09}] (1) If $P\!\subseteq\!\Sigma^\omega\!\times\! 2^\omega$, then we define 
$$\Game P\! :=\!\Big\{\sigma\!\in\!\Sigma^\omega\mid
\mbox{Player 1 has a winning strategy in the game } G(P_\sigma) \}$$
where $P_\sigma= \{\alpha\!\in\! 2^\omega\mid (\sigma ,\alpha )\!\in\! P\}$.\bigskip

\noindent (2) If $\bf\Gamma$ is a class of sets and $L\!\subseteq\!\Sigma^\omega$, then 
$L\!\in\!\Game {\bf\Gamma}$ if and only if there is 
$P\!\subseteq\!\Sigma^\omega\!\times\! 2^\omega$ in $\bf\Gamma$ such that $L\! =\!\Game P$.
\end{Deff}

 Note that one can replace $\Sigma^\om$ with the homeomorphic  space $T_\Si^\om$ in this definition, replacing also $P$ with a ${\bf\Gamma}$-subset of $T_\Si^\om \times 2^\om$. We obtain like this the definition of subsets of  $T_\Si^\om$ in the class $\Game ({\bf\Gamma})$. Note that we consider the alphabet $2=\{0, 1\}$  and not the finite alphabets with at least two letters $\Sigma$ in order to avoid troubles in the proof of Proposition \ref{tu} to come. This leads us to prove the following lemma.
 
\vfill\eject
 
\begin{Lem} \label{equi} We can find a continuous injection 
$\varphi\! :\!\Si^\om\!\rightarrow\! 2^\om$ and an open subset $O$ of $2^\om$ such that, for each subset $A$ of $\Si^\om$, the set $A'\! :=\!\varphi [A]\cup O$ satisfies the following property: Player 1 has a winning strategy in the game $G(A)$ if and only if Player 1 has a winning strategy in the game $G(A')$.\end{Lem} 

\noindent\bf Proof.\rm\ The proof is similar to that of Lemma 3.14 in \cite{Fin-2tape-extended}. We may assume that $\Sigma$ is a natural number and with a usual convention that $\Sigma=\{0, 1, \ldots , \Sigma -1\}$. Let $\varphi\! :\!\Si^\om\!\rightarrow\! 2^\om$ be defined by 
$\varphi (a_1a_2...)\! :=\! (11)^{a_1}0(11)^{a_2}0...$ Note that $\varphi$ is a continuous injection. We set $D_1\! :=\!\varphi [A]$, so that $D_1$ codes $A$. We also set 
$$\begin{array}{ll}
& D_2\! :=\!\{\alpha\!\in\! 2^\omega\mid\exists a_1,...,a_{2n}\!\in\!\Si~~\exists k\! <\!\Si\! -\! 1~~
(11)^{a_1}0...(11)^{a_{2n}}01^{2k+1}0\!\sqsubseteq\!\alpha\}\mbox{,}\cr
& D_3\! :=\!\{\alpha\!\in\! 2^\omega\mid\exists a_1,...,a_{2n+1}\!\in\!\Si ~~
(11)^{a_1}0...(11)^{a_{2n+1}}01^{2\Si -1}\!\sqsubseteq\!\alpha\} .
\end{array}$$
 As $D_2$ and $D_3$ are open, $O\! :=\! D_2\cup D_3$ is open too. The sets $D_2,D_3$ are the results of infinite plays where two players alternatively write a letter 0 or 1 and the infinite word written by the two players in $\omega$ steps is out of $K\! :=\!\varphi [\Si^\om ]$, due to the letters written by Player 2. More precisely, $D_2$ is the set of plays where Player 2 writes the 
$(2n\! +\! 1)$th letter 0 while it was Player 1's turn to do this. And $D_3$ is the set of plays where Player 2 does not write a letter 0 soon enough when it is his turn to do this.\bigskip

 If now the two players write alternatively a letter 0 or 1 in such a way that the infinite word written by  them in $\omega$ steps is in $K$ and of the form $(11)^{a_1}0(11)^{a_2}0...$, then the letters 0 have been written alternatively by Player 1 and by Player 2, and these letters 0 determine the natural numbers $a_i$. Thus the natural numbers $a_{2i+1}$ have been chosen by Player 1, and the natural numbers $a_{2i+2}$ have been chosen by Player 2 (for $i\!\in\!\omega$).\bigskip
 
 If Player 1 has a winning strategy $s$ in the game $G(A)$, then consider a play in the game 
$G(A')$. If the two players alternatively write a letter 0 or 1 and the infinite word $\alpha$ written by them in $\omega$ steps is in $K$ and of the form $(11)^{a_1}0(11)^{a_2}0...$, then the game is like a game where each player writes a letter in $\Si$ at each step of the play, and Player 1 can apply $s$ to ensure that $(a_i)_{i\geq 1}\!\in\! A$, which implies that 
$\varphi\big( (a_i)_{i\geq 1}\big)\!\in\! D_1\!\subseteq\! A'$, so Player 1 wins the play. If $\alpha$ is out of $K$ due to the letters written by Player 2, then $\alpha\!\in\! D_2\cup D_3\!\subseteq\! A'$, and Player 1 wins the play again. This shows that Player 1 has a winning strategy in $G(A')$.\bigskip

 If Player 1 has a winning strategy $s'$ in the game $G(A')$, then consider a play in the game 
$G(A)$ where Player 2 does not make the final word $\alpha$ in $D_2\cup D_3$. Player 1, following $s'$, must write letters so that the final word $\alpha$ belongs to $K$. Then the game is reduced to the game $G(A)$ in which the two players alternatively write letters $a_i$ in $\Si$. But Player 1 wins the game and this implies that Player 1 has a winning strategy in $G(A)$.
\hfill{$\square$}\bigskip 

 We now give the upper bound on the complexity of regular tree languages. 

\begin{The} \label{comptl} Let $\mathcal{A}$ be a Muller tree automaton. Then the tree language $L(\cal{A})$ is in the class $\Game (D_n({\bf \Sigma}_2^0))$, for some natural number $n\geq 1$.\end{The}

\proo Let $\mathcal{A}=(Q,\Si,\Delta, q_0, \mathcal{F})$ be a Muller tree automaton, where 
$(Q,\Si,\Delta, q_0)$ is a tree automaton and $\mathcal{F} \subseteq 2^Q$ is the collection of  designated state sets.\bigskip 

 The idea is to  use a game which was  considered by Gurevich and Harrington in \cite{GurevichH82}. For  $t\in  T_\Si^\om$ we consider the Gale-Stewart game $G(T_t)$, where the set $T_t \subseteq (\Delta\cup\{ l,r\})^\om$ is defined as follows. Intuitively, in the game $G(T_t)$, Player 1 writes transitions of the automaton 
 $\mathcal{A}$, i.e., letters of $\Delta$, and Player 2 writes letters $l$ or $r$, so that at the end of a play he has actually composed a path in the tree $t$.\bigskip

 At the first step, Player 1 chooses a transition $(q_0, t(\lambda), q, q') \in \Delta$. Next, Player 2 chooses $l$ or $r$. If Player 2 chooses $l$ this means he has chosen the left son of the root node of the tree and thus Player 1 has to choose a transition of the form $(q, t(l), q'', q''')$. If Player 2 chooses $r$ this means he has chosen the right son of the root node of the tree and thus Player 1 has to choose a transition of the form $(q', t(r), q'', q''')$. The game continues like this and Player 1 wins this game if the set of states (played by Player 1)  which appear infinitely often on the path chosen by Player 2 belongs to $\mathcal{F}$.\bigskip
 
 Formally, if $\alpha\!\in\!\Theta^\om$, then we define $(\alpha )_1,(\alpha )_2\!\in\!\Theta^\om$ by 
$(\alpha )_1 (k)\! :=\!\alpha (2k-1 )$ and  $(\alpha )_2 (k)\! :=\!\alpha (2k )$. $T_t$ is the union of 
$\Big\{\alpha\!\in\! (\Delta\cup\{ l,r\})^\om\mid\exists k\!\geq\! 1~(\alpha )_2(k)\!\notin\!\{ l,r\}\Big\}$ with\bigskip

\leftline{$\Big\{\alpha\!\in\! (\Delta\cup\{ l,r\})^\om\mid\forall k\!\geq\! 1~\Big( 
(\alpha )_1(k)\!\in\!\Delta\mbox{ and }(\alpha )_2(k)\!\in\!\{ l,r\}\mbox{ and }\alpha (1)(1)\! =\! q_0$}\smallskip

\rightline{$\mbox{and }(\alpha )_1(k)(2)\! =\! t\big( (\alpha )_2[k-1]\big)\mbox{ and }(\alpha )_1(k+1)(1)\! =\!\left\{\!\!\!\!\!\!\!
\begin{array}{ll}
& (\alpha )_1(k)(3)\mbox{ if }(\alpha )_2(k)\! =\! l\cr
& (\alpha )_1(k)(4)\mbox{ if }(\alpha )_2(k)\! =\! r
\end{array}
\right.\!\!\Big)$}\smallskip

\rightline{$\mbox{and }\{ q\!\in\! Q\mid\forall j\!\in\!\omega ~~\exists m\!\geq\! j
~~(\alpha )_1(m)(1)\! =\! q\}\!\in\! {\cal F}\Big\} .$}\bigskip

\noindent By definition of the Muller acceptance condition for the tree automaton $\mathcal{A}$, 
$$t \in L({\cal A})\Leftrightarrow\mbox{  Player 1 has a winning strategy in the game }G (T_t).$$ 
Indeed, if $t \in L({\cal A})$, then there is a witness $\rho\in T_Q^\om$ and we can define a strategy 
$\tilde\rho$ for Player 1 by $\tilde\rho (x)\! :=\!\big(\rho (x), t(x),\rho (xl),\rho (xr)\big)$, and 
$\tilde\rho$ is winning for Player 1. Conversely, if $\tilde\rho$ is a winning strategy for Player 1, then the formula $\rho (x)\! :=\!\tilde\rho (x)(1)$ defines a witness for the fact that $t \in L({\cal A})$.\bigskip 

 We apply Lemma \ref{equi} to the finite alphabet $\Delta \cup \{l,r\}$, which gives a continuous injection $\varphi\! :\! (\Delta \cup \{l,r\} )^\om\!\rightarrow\! 2^\om$ and an open subset $O$ of 
$2^\omega$. Then we set
$$P\! :=\!\big\{ (t,\alpha ) \in T_\Si^\om \times 2^\om\mid\alpha\in\varphi [T_t]\cup O\big\}\mbox{,}$$
so that $t \in L({\cal A})\Leftrightarrow\mbox{  Player 1 has a winning strategy in the game }G (P_t)
\Leftrightarrow t\!\in\!\Game P$.

\vfill\eject

 It remains to see that $P$ is in $D_n({\bf \Sigma}_2^0)$ for some natural number $n\geq 1$. If 
$B\! :=\! D_n[(A_p)_{p<n}]$ is in $D_n({\bf \Sigma}_2^0)$ and $C$ is a closed set, then 
$B\cap C\! =\! D_n[(A_p\cap C)_{p<n}]$ is also in $D_n({\bf \Sigma}_2^0)$. If $U$ is an open set, then $B\cup U\! =\! D_n[(A_p\cup U)_{p<n}]$ if $n$ is odd, and 
$B\cup U\! =\! D_n[(A_0\!\setminus\! U,A_1\cup U,...,A_{n-1}\cup U)]$ if $n$ is even, so that 
$B\cup U$ is also in $D_n({\bf \Sigma}_2^0)$. This shows that the class $D_n({\bf \Sigma}_2^0)$ is closed under intersections with a closed set and unions with an open set. As it is also closed under images by homeomorphisms and $\varphi$ is a homeomorphism onto its compact range, it is enough to show that $T$ is in $D_n({\bf \Sigma}_2^0)$ for some natural number $n\geq 1$. This is known to be equivalent to the statement ``$T$ is a boolean combination of ${\bf \Sigma}_2^0$-sets", and this follows from the definition of the  Muller acceptance condition of the tree automaton 
$\mathcal{A}$. Indeed,\bigskip

- the condition ``$\exists k\!\geq\! 1~(\alpha )_2(k)\!\notin\!\{ l,r\}$" is open,\bigskip

- the condition involving $k$ and the big parentheses in the definition of $T_t$ is closed,\bigskip

- the condition involving $\cal F$ says that there is $F$ in the finite set $\cal F$ such that, for each $q$ in the finite set $Q$, either $q\!\in\! F$ and $\forall j\!\in\!\omega ~~\exists m\!\geq\! j~~
(\alpha )_1(m)(1)\! =\! q$ (this is a ${\bf\Pi}^0_2$ condition), or $q\!\notin\! F$ and 
$\exists j\!\in\!\omega ~~\forall m\!\geq\! j~~(\alpha )_1(m)(1)\!\not=\! q$ (this is a ${\bf\Sigma}^0_2$ condition). This finishes the proof.\hfill{$\square$}

\section{$\!\!\!\!\!\!\!$ The  Upper Bound  is much better than ${\bf \Delta}_2^1$}\indent

 In this section, we are going to show that the upper bound given in the preceding section is actually much better than the usual one, ${\bf \Delta}_2^1$.\bigskip 
 
\noindent $\bullet$ We first define the Wadge hierarchy, which is a great refinement of the Borel hierarchy, and is  defined via reductions by continuous functions, \cite{Duparc01,Wadge83}. 

\begin{Deff}[Wadge \cite{Wadge83}] Let $L\subseteq\Si^\om$ and $L'\subseteq\Theta^\om$. We say that $L$ and $L'$ are {\bf Wadge equivalent} if $L\leq _W L'$ and $L'\leq _W L$. This will be denoted by $L\equiv_W L'$. We shall also say that $L<_W L'$ iff $L\leq _W L'$ and 
$L'\not\leq _W L$. We say that $L$ is {\bf self dual} if $L\equiv_W\neg L$ ($\neg L$ is the complement of $L$), otherwise we say that $L$ is {\bf non self dual}.\end{Deff}

 The relation $\leq _W $  is reflexive and transitive, and $\equiv_W $ is an equivalence relation. The equivalence classes of $\equiv_W $ are called {\bf Wadge degrees}. The 
{\bf Wadge hierarchy} $WH$ is the class of Borel subsets of a space  $\Si^\om$, equipped with 
$\leq _W $ and $\equiv_W $. For $L\subseteq\Si^\om$ and $L'\subseteq\Theta^\om$, if 
$L\leq _W L'$ and $L=f^{-1}(L')$, where $f$ is a continuous function from $\Si^\om$ into 
$\Theta^\om$, then $f$ is called a {\bf continuous reduction} of $L$ to $L'$. Intuitively, it means that $L$ is less complicated than $L'$, because in order to check whether $\sigma\in L$, it suffices to check whether $f(\sigma )\in L'$ (where $f$ is a continuous function). Hence the Wadge degree of an\ol~is a measure of its topological complexity.

\vfill\eject
  
 Note that in the above definition, we consider that a subset $L\subseteq\Si^\om$ is given
together with the alphabet $\Si$. This is important as it is shown by the following simple example. 
Let $L_1=\{0, 1\}^\om \subseteq \{0, 1\}^\om$ and $L_2=\{0, 1\}^\om \subseteq \{0, 1, 2\}^\om$. So the languages $L_1$ and $L_2$ are equal but considered over the different alphabets 
$\Si_1=\{0, 1\}$ and $\Si_2=\{0, 1, 2\}$. It turns out that $L_1 <_W L_2$. In fact $L_1$ is open 
{\it and } closed in $\Si_1^\om$ while $L_2$ is closed but not open in $\Si_2^\om$.\bigskip 

 We can now define the Wadge class of a set $L$.

\begin{Deff} Let $L$ be a subset of $\Si^\om$. The {\bf Wadge class} of $L$ is
$$[L]=\{ L'\mid L'\subseteq\Theta^\om\mbox{ for a finite alphabet }\Theta\mbox{  and  }
L'\leq _W L\}.$$\end{Deff}

 Recall that each Borel class ${\bf \Si}^0_\xi$ and ${\bf \Pi}^0_\xi$ is a Wadge class. It follows from the study of the Wadge hierarchy that a set $L\subseteq X^\om$ is ${\bf \Si}^0_\xi$ (respectively, 
${\bf \Pi}^0_\xi$)-complete iff it is in ${\bf \Si}^0_\xi$ but not in ${\bf \Pi}^0_\xi$ (respectively, in 
${\bf \Pi}^0_\xi$ but not in ${\bf \Si}^0_\xi$).\bigskip

 Using the notion of a Wadge game, and the fact that the determinacy of Wadge games follows from  Martin's Theorem stating  that every Gale-Stewart Game $G(B)$, with $B$ a Borel set, is determined, see \cite{Kechris94}, Wadge proved the following result. 

\begin{The} [Wadge]\label{wh} Up to complement and $\equiv _W$, the class of Borel subsets of the spaces $\Si^\om$ is a well ordered hierarchy. We can find an ordinal $|WH|$, called the length of the hierarchy, and a map $d_W^0$ from $WH$ onto $|WH|\setminus\{0\}$, such that, for all 
$L, L' \subseteq\Si^\om$,\smallskip 

- $d_W^0 L < d_W^0 L' \Leftrightarrow L<_W L' $,\smallskip

- $d_W^0 L = d_W^0 L' \Leftrightarrow [ L\equiv_W L' $ or $L\equiv_W\neg L']$.\end{The}

\noindent $\bullet$ Let $\omega_1$ be the {\bf first uncountable ordinal}. The Wadge hierarchy of Borel sets of finite rank has length $^1\varepsilon_0$ where $^1\varepsilon_0$ is the limit of the ordinals $\alpha_n$ defined by $\alpha_1=\om_1$ and $\alpha_{n+1}=\om_1^{\alpha_n}$ if $n$ is a natural number. Then $^1\varepsilon_0$ is the first fixed point of the ordinal exponentiation of base $\om_1$. The length of the Wadge hierarchy of Borel sets in 
${\bf \Delta}^0_\om ={\bf\Si}^0_\om\cap {\bf\Pi}^0_\om$ is the $\om_1^{th}$ fixed point of the ordinal exponentiation of base $\om_1$, which is a much larger ordinal. The length of the whole Wadge hierarchy is described in \cite{Wadge83,Duparc01}, and uses the Veblen functions. To recall the definition of these functions, we  need the notion of cofinality of an ordinal  which may be found in \cite{Jech} and which we briefly recall now.

\begin{Deff} Let $\zeta$ be a limit ordinal. The {\bf cofinality} of $\zeta$, denoted by $cof(\zeta)$,
is the least ordinal $\beta$ such that there exists a strictly increasing sequence of ordinals
$(\zeta_i)_{i<\beta}$, of length $\beta$, such that $\sup_{i< \beta} \zeta_i = \zeta$ and 
$\zeta_i < \zeta$ for each $i< \beta$. This definition is usually extended to 0 and to the successor ordinals: $cof(0)=0$ and $cof(\zeta +1)=1 \mbox{ for every ordinal  } \zeta$.\end{Deff}

 The cofinality of a  limit ordinal  is always a limit ordinal with $\om\leq cof(\zeta)\leq\zeta$. The ordinal $cof(\zeta)$ is in fact a cardinal (see \cite{Jech}). If the cofinality of a limit ordinal $\zeta$ is 
$\leq \om_1$, then only the following cases may happen: $cof(\zeta )=\om$ or $cof(\zeta )=\om_1$. In the sequel we do not need to consider cofinalities which are larger than $\om_1$. In the sequel, $\om_2$ willl be the first ordinal of cardinality greater than $\aleph_1$, the cardinal of $\om_1$.  Note that each ordinal $\zeta <\om_2$ has cofinality smaller than or equal to $\om_1$. 

\begin{Deff} The {\bf Veblen hierarchy} $(V_\xi )_{\xi < \om_1}$ of functions from
 $\om_2 \setminus \{0\}$ into itself is defined as follows. $V_0$ is the function that enumerates ordinals of cofinality $\om_1$ or 1 that are closed under ordinal addition, i.e., 
$V_0(1)=1$, $V_0(\zeta + 1) = V_0(\zeta ) \cdot \om_1$, $V_0(\zeta ) = \om_1^{\zeta}$  when 
$cof(\zeta )=\om_1$, $V_0(\zeta ) = \om_1^{\zeta + 1}$ when $cof(\zeta )=\om$. 
For $\xi > 0$, $V_\xi$ is the function that enumerates ordinals of cofinality $\om_1$ or 1 that are closed under each function $V_\eta$ for any $\eta < \xi$.\end{Deff}

 We can now describe the length of the whole Wadge hierarchy of Borel sets on a Cantor space 
$\Sio$ or $T_\Si^\om$. It is the ordinal $\sup_{\xi < \om_1} V_\xi (2)$. This is really a huge ordinal, with regard to the $\om_1^{th}$ fixed point of the ordinal exponentiation of base $\om_1$, which is the length of the  Wadge hierarchy of Borel sets in ${\bf \Delta^0_\om}$.\bigskip

\noindent $\bullet$ We now recall the notion of a universal set which will be useful in the sequel. 

\begin{Deff} Let $\bf\Gamma$ be a class of sets. We say that 
${\cal U}\!\subseteq\! 2^\omega\!\times\!\Sigma^\omega$ is \bf universal\it\ for the $\bf\Gamma$ subsets of $\Sigma^\omega$ if $\cal U$ is in $\bf\Gamma$, and for each 
$A\!\subseteq\!\Sigma^\omega$ in $\bf\Gamma$ there is $\beta\!\in\! 2^\omega$ such that 
$A$ is the vertical section 
${\cal U}_\beta\! :=\!\{\alpha\!\in\!\Sigma^\omega\mid (\beta ,\alpha )\!\in\! {\cal U}\}$ of $\cal U$ at 
$\beta$.\end{Deff}

 The following result is mentioned in \cite{Louveau-Saint-Raymond}.
 
\begin{The} \label{sdu} Let $\bf\Gamma$ be the Wadge class of a non self-dual Borel subset of 
$2^\omega$. Then there is a universal set for the $\bf\Gamma$ subsets of $\Sigma^\omega$.
\end{The}

\noindent\bf Proof.\rm\ Let $A,B\!\subseteq\! 2^\omega$. We consider the {\bf Wadge game} 
$G(A,B)$ associated with $A$ 	and $B$. It is the Gale-Stewart game defined by 
$$\mbox{Player 2 wins the play}~\Leftrightarrow ~
(a_1a_2...\!\in\! A\Leftrightarrow b_1b_2...\!\in\! B).$$ 
If $s$ is a strategy for Player 2 and $\alpha\!\in\! 2^\omega$, then we denote by $s\! *\!\alpha$ the element $b_1b_2...$ of $2^\omega$ given by the answers of Player 2 in the play where Player 1 plays $\alpha\! :=\! a_1a_2...$ and Player 2 follows $s$. Formally, 
$(s\! *\!\alpha )(1)\! :=\! b_1\! :=\! s(a_1)$ and, inductively, 
$(s\! *\!\alpha )(k)\! :=\! b_k\! :=\! s(a_1b_1a_2...b_{k-1}a_k)$ if $k\!\geq\! 2$. Notice that a strategy $s$ for Player 2 
is an element of $2^{(2^2)^*2}$. 
As $(2^2)^*2$ is countable, we can identify $2^{(2^2)^*2}$ with the Cantor space $2^\omega$. Note that the map 
$e\! :\! (s,\alpha )\!\mapsto\! s\! *\!\alpha$ is continuous from $2^{(2^2)^*2}\!\times\! 2^\omega$ into 
$2^\omega$.\bigskip

 Theorem 2.7 and Remark 4 after the proof of Theorem 3.1 in \cite{Louveau-Saint-Raymond} provide a subset $L$ of $2^\omega$ in $\bf\Gamma$ such that Player 2 has a winning strategy in the game $G(A,L)$ for each subset $A$ of $2^\omega$ in $\bf\Gamma$. We set 
${\cal V}\! :=\!\{ (s,\alpha )\!\in\! 2^{(2^2)^*2}\!\times\! 2^\omega\mid s\! *\!\alpha\!\in\! L\}$. As $e$ is continuous, $\cal V$ is in $\bf\Gamma$. If $A\!\subseteq\! 2^\omega$ is in $\bf\Gamma$, then Player 2 wins the Wadge game $G(A,L)$, which gives $s$ in $2^{(2^2)^*2}$ such that 
$A\! =\! {\cal V}_s$. This shows that we can consider $\cal V$ as universal for the $\bf\Gamma$ subsets of $2^\om$, up to identification.\bigskip

 Let $\Sigma$ be a finite alphabet, and $\psi\! :\!\Si^\om\!\rightarrow\! 2^\omega$ be a homeomorphism. We set ${\cal U}\! :=\! (\mbox{Id}_{2^\omega}\!\times\!\psi )^{-1}({\cal V})$, and it is routine to check that $\cal U$ is universal for the $\bf\Gamma$ subsets of $\Si^\om$.
 \hfill{$\square$}

\begin{Pro} \label{tu} Let $\bf\Gamma$ be a class of sets. If there is a universal for the $\bf\Gamma$ subsets of $\Sigma^\omega\!\times\! 2^\omega$, then there is a universal for the 
$\Game {\bf\Gamma}$ subsets of $\Sigma^\omega$.\end{Pro} 

\noindent\bf Proof.\rm\ Let $\cal U$ be a universal for the $\bf\Gamma$ subsets of 
$\Sigma^\omega\!\times\! 2^\omega$. We set ${\cal V}\! :=\!\Game {\cal U}$. Then 
${\cal V}\!\subseteq\! 2^\omega\!\times\!\Sigma^\omega$ is in $\Game {\bf\Gamma}$. Let 
$A\!\subseteq\!\Sigma^\omega$ be in $\Game {\bf\Gamma}$, and 
$P\!\subseteq\!\Sigma^\omega\!\times\! 2^\omega$ in $\bf\Gamma$ such that 
$A\! =\!\Game P$. Then there is $\beta\!\in\! 2^\omega$ with $P\! =\! {\cal U}_\beta$. It remains
 to note that $A\! =\! {\cal V}_\beta$.\hfill{$\square$}

\begin{Cor} \label{ug} Let $\bf\Gamma$ be the Wadge class of a non self-dual Borel subset of 
$2^\omega$. Then there is a universal for the $\Game {\bf\Gamma}$ subsets of $\Sigma^\omega$.\end{Cor} 

\noindent\bf Proof.\rm\ Theorem \ref{sdu} gives a universal for the $\bf\Gamma$ subsets of 
$\Sigma^\omega\!\times\! 2^\omega$, and Proposition \ref{tu} gives the result.\hfill{$\square$}\bigskip

\noindent $\bullet$ We now turn to the key result of embeddability of the Wadge hierarchy into that obtained with the game quantifier.

\begin{Lem} \label{classe} Let $\bf\Gamma$ be a class of sets closed under continuous pre-images. Then $\Game {\bf\Gamma}$ is also closed under continuous pre-images.\end{Lem} 

\noindent\bf Proof.\rm\ Let $L\!\subseteq\!\Sigma^\omega$ in $\Game {\bf\Gamma}$, 
$P\!\subseteq\!\Sigma^\omega\!\times\! 2^\omega$ in $\bf\Gamma$ such that 
$L\! =\!\Game P$, and $f\! :\!\Theta^\omega\!\rightarrow\!\Sigma^\omega$ be continuous. As 
$\bf\Gamma$ is closed under continuous pre-images, the set 
$$Q\! :=\!
\{ (\delta ,\alpha )\!\in\!\Theta^\omega\!\times\! 2^\omega\mid\big( f(\delta ),\alpha\big)\!\in\! P\}$$ 
is in $\bf\Gamma$. Note that, for each $\delta\!\in\!\Theta^\omega$, 
$$\begin{array}{ll}
\delta\!\in\! f^{-1}(L)\!\!\!\!
& \Leftrightarrow f(\delta )\!\in\!\Game P\cr
& \Leftrightarrow\mbox{Player 1 has a winning strategy in the game }\cr
& \hfill{G(\big\{\alpha\!\in\! 2^\omega\vert\big( f(\delta ),\alpha\big)\!\in\! P\big\} )}\cr
& \Leftrightarrow\mbox{Player 1 has a winning strategy in the game }
G(\{\alpha\!\in\! 2^\omega\vert (\delta ,\alpha )\!\in\! Q\} )\cr
& \Leftrightarrow\delta\!\in\!\Game Q.
\end{array}$$
Thus $f^{-1}(L)\!\subseteq\!\Theta^\omega$ is in $\Game {\bf\Gamma}$.\hfill{$\square$}

\begin{Cor}\label{embedding} 
\label{strict} The map ${\bf\Gamma}\!\mapsto\!\Game {\bf\Gamma}$, defined on the collection 
of Wadge classes of a non self-dual Borel subset of $2^\omega$, is strictly increasing for the inclusion.
\end{Cor} 

\noindent\bf Proof.\rm\ Let $L,L'$ be non self-dual Borel subsets of $2^\omega$, and 
${\bf\Gamma},{\bf\Gamma}'$ be the Wadge classes of $L,L'$ respectively. We assume that 
${\bf\Gamma}\!\subsetneqq\! {\bf\Gamma}'$, so that $L<_WL'$. By Wadge's lemma, 
$\neg L\leq_WL'$, so that ${\bf\Gamma}$ and $\check {\bf\Gamma}\! :=\!\{\neg A\mid A\!\in\! {\bf\Gamma}\}$ 
are contained in ${\bf\Gamma}'$ (see 21.14 in [Kec95]). This implies that 
$\Game {\bf\Gamma}\!\subseteq\!\Game {\bf\Gamma}'$, and in fact 
$\Game {\bf\Gamma}\cup\Game\check {\bf\Gamma}\!\subseteq\!\Game {\bf\Gamma}'$.\bigskip

 Assume, towards a contradiction, that $\Game {\bf\Gamma}\! =\!\Game {\bf\Gamma}'$. Corollary 
\ref{ug} gives a universal $\cal U$ for the $\Game {\bf\Gamma}'$ subsets of $2^\omega$. We set 
$A\! :=\!\{\alpha\!\in\! 2^\omega\mid (\alpha ,\alpha )\!\notin\! {\cal U}\}$. As $\cal U$ is in 
$\Game {\bf\Gamma}'$, it is also in $\Game {\bf\Gamma}$. This gives 
$P\!\subseteq\! (2^\omega )^3$ in $\bf\Gamma$ with ${\cal U}\! =\!\Game P$. As $L$ is Borel, 20.5 in [Kec95] implies that 
$$\begin{array}{ll}
(\beta ,\gamma )\!\notin\! {\cal U}\!\!\!\!
& \Leftrightarrow\neg\big(\mbox{Player 1 has a winning strategy in }
G(\{\alpha\!\in\! 2^\omega\mid (\beta ,\gamma ,\alpha )\!\in\! P\} )\big)\cr
& \Leftrightarrow\mbox{Player 2 has a winning strategy in }
G(\{\alpha\!\in\! 2^\omega\mid (\beta ,\gamma ,\alpha )\!\notin\! P\} ).
\end{array}$$
By 6D.1 in [Mos09], $\neg {\cal U}\!\in\!\Game\check {\bf\Gamma}\!\subseteq\!\Game {\bf\Gamma}'$. As the map $\alpha\!\mapsto\! (\alpha ,\alpha )$ is continuous, $A\!\in\!\Game {\bf\Gamma}'$, by Lemma \ref{classe}. This gives $\beta\!\in\!\omega^\omega$ such that $A\! =\! {\cal U}_\beta$. Thus 
$$(\beta ,\beta )\!\in\! {\cal U}\Leftrightarrow\beta\!\in\! A\Leftrightarrow 
(\beta ,\beta )\!\notin\! {\cal U}\mbox{,}$$ 
which is absurd.\hfill{$\square$}

\begin{Lem} \label{inclusion} Let $\bf\Gamma$ be a class of Borel sets. Then 
$\Game {\bf\Gamma}$ is contained in ${\bf\Delta}^1_2$.\end{Lem} 

\noindent\bf Proof.\rm\ Let $L\!\subseteq\!\Sigma^\omega$ be in $\Game {\bf\Gamma}$, and 
$P\!\subseteq\!\Sigma^\omega\!\times\! 2^\omega$ be in $\bf\Gamma$ such that $L\! =\!\Game P$. If 
$\sigma ,\tau$ are strategies for Players 1 and 2 respectively, then we denote by $\sigma *\tau$ the element of $2^\omega$ obtained when Player 1 follows $\sigma$ and Player 2 follows $\tau$. As $P$ is Borel, $P_\beta\! :=\!\{\alpha\!\in\! 2^\omega\mid (\beta ,\alpha )\!\in\! P\}$ is Borel for each 
$\beta\!\in\!\Sigma^\omega$. By 20.5 in [Kec95], the game defined by $P_\beta$ is determined. This implies that $\beta\!\in\! L$ iff
$$\begin{array}{ll}
\beta\!\in\!\Game P\!\!\!\!
& \Leftrightarrow\mbox{Player 1 has a winning strategy in the game }
G(\{\alpha\!\in\! 2^\omega\mid (\beta ,\alpha )\!\in\! P\} )\cr
& \Leftrightarrow\exists\sigma ~\forall\tau ~(\beta ,\sigma *\tau )\!\in\! P
\Leftrightarrow\forall\tau ~\exists\sigma ~(\beta ,\sigma *\tau )\!\in\! P.
\end{array}$$
Thus $L\!\in\!{\bf\Delta}^1_2$.\hfill{$\square$}\bigskip

 The classes $D_n(\boratwo )$ appearing in Theorem \ref{comptl} are contained in $\borathree$, 
which is a small part of the class of Borel sets. By Theorem \ref{comptl} and Corollary \ref{strict}, 
the tree languages of Muller tree automata are in the class $\Game\borathree$. By 22.4 in [Kec95], the sequence $({\bf \Si}^0_\xi )_{\xi <\omega_1}$ is strictly increasing for the inclusion. By Corollary \ref{strict}, the sequence $(\Game {\bf \Si}^0_\xi )_{\xi <\omega_1}$ is also strictly increasing for the inclusion. By Lemma \ref{inclusion}, the elements of the latter sequence are contained in the class 
${\bf\Delta}^1_2$. This shows already that the increasing sequence 
$\big(\Game (D_n(\boratwo ))\big)_{n\geq 1}$ is only a small part of the hierarchy of 
${\bf\Delta}^1_2$ sets.

\vfill\eject

 This is actually emphasized by considering the Wadge hierarchy of non self-dual Borel subsets of $2^\omega$. By Corollary \ref{embedding} we know that this hierarchy can be embedded into a hierarchy of classes $\Game {\bf\Gamma}$, included into ${\bf\Delta}^1_2$. On the other hand the Borel class $\boratwo$ is known to be a non self-dual Wadge class and the Wadge degree (in the sense of Theorem \ref{wh}) of any $\boratwo$-complete set is equal to $\om_1$. Moreover, for each natural number $n\geq 1$,  the class $D_n(\boratwo )$ is also a non self-dual Wadge class and the Wadge degree of any  $D_n(\boratwo )$-complete is equal to $(\om_1)^n$, \cite{Wadge83,Duparc01}. Therefore, we see that this ordinal is actually much smaller than the first fixed point of the ordinal exponentiation of base $\om_1$ (the length of the Wadge hierarchy of Borel sets of finite ranks), than the $\om_1^{th}$ fixed point of the ordinal exponentiation of base 
$\om_1$ (the length of the Wadge hierarchy of Borel sets in ${\bf \Delta^0_\om}$),  and a fortiori than the ordinal 
$$\sup_{\xi < \om_1} V_\xi (2)\mbox{,}$$
which is the length of the Wadge hierarchy of (non self-dual) Borel subsets of the Baire space or the Cantor space. In conclusion the upper bound we gave is much smaller than ${\bf\Delta}^1_2$, as summarized in the following theorem. 

\begin{The}\label{hierarchy} 
The hierarchy of classes  $\Game {\bf\Gamma}$, for non self-dual Borel Wadge classes ${\bf\Gamma}$ of $2^\omega$, 
is a hierarchy of length $\sup_{\xi < \om_1} V_\xi (2)$   included in the class ${\bf\Delta}^1_2$,  while the regular tree languages are contained 
in the first  $(\om_1)^\omega$ levels of this hierarchy. 
\end{The}

\noindent $\bullet$ We also note that regular tree languages are in the class 
provably-$\Delta_2^1$ (see \cite[page 180]{Kanamori}). This is essentially proved in 
\cite{HjorthKMN08} from Rabin's Theorem \cite{Rabin69}. Thus all regular tree languages 
have the Baire property (see \cite[page 180]{Kanamori}). This gives an answer to a question of 
the third author in \cite{Simonnet92}.\bigskip

 As noticed by the referee of this paper, this also follows from \cite{GogaczMMS14}, since all R-sets are Baire-measurable as well as Lebesgue measurable and it is proved in \cite{GogaczMMS14} that regular tree languages are R-sets. 

\section{$\!\!\!\!\!\!\!$ Concluding remarks}\indent

 We gave an upper bound on the topological complexity of recognizable tree languages and showed, using an embedding of the Wadge hierarchy of non self-dual Borel sets of $2^\om$, that this upper bound is actually much smaller than ${\bf\Delta}^1_2$.\bigskip
 
 The anonymous referee of this paper indicated us that since the first appearance in 1992  of the above Theorem \ref{comptl} in the PhD Thesis of the third author of this paper, quite a few authors made various observations in a few papers that we now mention.\bigskip
 
  In J. Bradfield's paper \cite{Bradfield03}, and in \cite{BradfieldDQ05} by J. Bradfield, J. Duparc and S. Quickert, a link between the game quantifier and the $\mu$-calculus is described, and Corollary 11 of \cite{Bradfield03} explicitely states an upper bound in a style similar to the one presented in our paper. In \cite{MichalewskiN12}, D. Niwinski and H. Michalewski are also interested in the problem of finding upper bounds for the class of regular tree languages. Using a method developed by J. Saint Raymond in \cite{SRaymond}, they prove that the game tree language 
$W_{1, 3}$ is complete for the class of  ${\bf \Sigma}_1^1$-inductive sets. The ${\bf \Sigma}_1^1$-inductive sets are known to contain more complex sets than the $\sigma$-algebra generated by the analytic sets. Moreover, the referee indicated us that one can actually verify that the language given in S. Hummel's paper \cite{Hummel12} is reducible to $W_{1, 3}$ and is not complete for the class of  ${\bf \Sigma}_1^1$-inductive sets. In another recent paper \cite{GogaczMMS14} 
T. Gogacz, H.  Michalewski, M. Mio and  M. Skrzypczak show a one-to-one correspondence between the levels of the hierarchy of Kolmogorov R-sets and parity index of regular languages which extends the theorem from \cite{MichalewskiN12}. Since due to a theorem of Burgess the R-sets are known to be in correspondence with the game quantifier, on a technical level this covers  Theorem \ref{comptl}.  We also notice that another  estimation of the kind of Theorem 
\ref{hierarchy} is present in \cite{GogaczMMS14}, where it is shown that regular languages of infnite trees occupy  exactly the first $\om$-levels of Kolmogorov's hierarchy; and on the other hand it is known that there exist $\om_1$ levels in this hierarchy, all of them contained in the class 
${\bf\Delta}^1_2$. Theorem  \ref{hierarchy} may then be seen as a generalization of the estimation following from \cite{GogaczMMS14}.\bigskip

 We now state some important open questions.\bigskip

\noindent $\bullet$ A difficult problem in the study of the topological complexity of recognizable tree languages is to determine the Wadge hierarchy of tree languages accepted by non deterministic Muller or Rabin tree automata. A subquestion is to know whether there is a regular language of infinite trees which is a Borel set of {\it infinite rank}.\bigskip

\noindent $\bullet$ A related problem is to study the determinacy of Wadge games between tree automata; this would be a first step towards the possibility of using these games to determine the topological complexity of a regular tree language. Note that in the case of one-counter B\"uchi automata it has been shown that the determinacy of such games needs some large cardinal assumption (see \cite{Fin13-JSL}).\bigskip

 We now state some important open questions which are more related to the computer science roots of regular languages since they also involve decidability questions.\bigskip  

\noindent $\bullet$  While many questions about {\it deterministic} regular languages of infinite trees have been shown to be decidable \cite{ADMN},  many corresponding questions about {\it non-deterministic} regular languages of infinite trees are still open.\bigskip

 Concerning the decidability of the topological complexity of regular languages, it has been shown recently by M. Bojanczyk and T. Place in \cite{BojanczykP12} that one can decide whether a regular language accepted by a given tree automaton is a boolean combination of open sets. This result has been extended by A. Facchini and H. Michalewski in \cite{FacchiniM14}, where the authors prove that one can decide whether a regular tree language is in the class 
${\bf \Delta}_2^0$. The question is still open for the other levels of the Borel hierarchy, or whether a regular tree language is Borel, analytic, coanalytic, or in any class $\Game {\bf\Gamma}$ present in this paper.  \bigskip  

\noindent {\bf Acknowledgement.} We wish to thank   the  anonymous referee for very useful comments on a preliminary version of this paper which lead to a great improvement of the paper, and for indicating us some related studies that are now present in the bibliography. 

\nocite{Simonnet92, Kechris94}

\nocite{Moschovakis09,Murlak-LMCS} 
\nocite{PerrinPin,LescowThomas,Rabin69,CS07,Thomas97,Fink-Sim-trees,Louveau-Saint-Raymond}
\nocite{GogaczMMS14,MichalewskiN12,SRaymond,Bradfield03,BradfieldDQ05,BojanczykP12,FacchiniM14}

\bibliographystyle{alpha}
\bibliography{mabiblio,mabiblio-trees}

\end{document}